\documentclass[prl,aps,showpacs,groupedaddress,superscriptaddress,twocolumn,toc=flat,colorinlistoftodos]{revtex4-1}

\usepackage{braket}
\usepackage{amsmath,amssymb}
\usepackage{graphicx}
\usepackage[dvipsnames]{xcolor}
\usepackage[caption=false]{subfig}
\usepackage[colorlinks=true,linkcolor=blue,urlcolor=blue,citecolor=blue]{hyperref}

\usepackage{nicefrac}

\usepackage{todonotes}
\usepackage{blindtext}

\begin{document}
\title{Ferromagnetism and Skyrmions in the Hofstadter-Fermi-Hubbard Model}


\author{F. A. Palm}
\affiliation{Department of Physics and Arnold Sommerfeld Center for Theoretical Physics (ASC), Ludwig-Maximilians-Universit\"at M\"unchen, Theresienstr. 37, D-80333 M\"unchen, Germany}
\affiliation{Munich Center for Quantum Science and Technology (MCQST), Schellingstr. 4, D-80799 M\"unchen, Germany}

\author{M. Kurttutan}
\affiliation{Department of Physics and Arnold Sommerfeld Center for Theoretical Physics (ASC), Ludwig-Maximilians-Universit\"at M\"unchen, Theresienstr. 37, D-80333 M\"unchen, Germany}
\affiliation{Munich Center for Quantum Science and Technology (MCQST), Schellingstr. 4, D-80799 M\"unchen, Germany}
\affiliation{Department of Physics, Freie Universit\"at Berlin, Arnimallee 14, D-14195 Berlin, Germany}

\author{A. Bohrdt}
\affiliation{Department of Physics, Harvard University, Cambridge, MA 02138, USA}
\affiliation{ITAMP, Harvard-Smithsonian Center for Astrophysics, Cambridge, MA 02138, USA}

\author{U. Schollw\"ock}
\affiliation{Department of Physics and Arnold Sommerfeld Center for Theoretical Physics (ASC), Ludwig-Maximilians-Universit\"at M\"unchen, Theresienstr. 37, D-80333 M\"unchen, Germany}
\affiliation{Munich Center for Quantum Science and Technology (MCQST), Schellingstr. 4, D-80799 M\"unchen, Germany}

\author{F. Grusdt}
\affiliation{Department of Physics and Arnold Sommerfeld Center for Theoretical Physics (ASC), Ludwig-Maximilians-Universit\"at M\"unchen, Theresienstr. 37, D-80333 M\"unchen, Germany}
\affiliation{Munich Center for Quantum Science and Technology (MCQST), Schellingstr. 4, D-80799 M\"unchen, Germany}

\date{\today}

\begin{abstract}
Strongly interacting fermionic systems host a variety of interesting quantum many-body states with exotic excitations.
For instance, the interplay of strong interactions and the Pauli exclusion principle can lead to Stoner ferromagnetism, but the fate of this state remains unclear when kinetic terms are added.
While in many lattice models the fermions' dispersion results in delocalization and destabilization of the ferromagnet, flat bands can restore strong interaction effects and ferromagnetic correlations.
To reveal this interplay, here we propose to study the Hofstadter-Fermi-Hubbard model using ultracold atoms.
We demonstrate, by performing large-scale DMRG simulations, that this model exhibits a lattice analog of the quantum Hall ferromagnet at magnetic filling factor $\nu=1$.
We reveal the nature of the low energy spin-singlet states around $\nu\approx1$ and find that they host quasi-particles and quasi-holes exhibiting spin-spin correlations reminiscent of skyrmions.
Finally, we predict the breakdown of flat-band ferromagnetism at large fields.
%
%
Our work paves the way towards experimental studies of lattice quantum Hall ferromagnetism, including prospects to study many-body states of interacting skyrmions and explore the relation to high-$T_{\rm c}$ superconductivity.
\end{abstract}

\maketitle

\paragraph{Introduction.---}
The emergence of ferromagnetism in strongly correlated systems is a phenomenon well-known in condensed matter physics and subject of ongoing research.
In particular, the interplay of strong repulsion and the Pauli exclusion principle in fermionic systems can give rise to the phenomenon known as Stoner ferromagnetism.
Due to enhanced interaction effects in flat bands, quantum Hall (QH) systems provide a powerful platform to study this mechanism.

Indeed, at magnetic filling factor $\nu=1$, the ground state of the two-dimensional electron gas was predicted to be ferromagnetically ordered~\cite{Yang1994,Ezawa2009}.
Furthermore, the low-lying, charged quasi-particle excitations around this QH~ferromagnet have been predicted to exhibit exotic spin textures known as skyrmions.
These are characterized by local ferromagnetic correlations while being in an overall spin-singlet state~\cite{Rezayi1987,Haldane1988,Lee1990,Rezayi1991,Sondhi1993,Fertig1994,Moon1995,MacDonald1996}.
The QH~ferromagnet and its skyrmionic excitations were observed in solid state experiments~\cite{Barrett1995,Schmeller1995,Aifer1996,Manfra1997,Townsley2005,Bryja2006,Lupatini2020} and attracted considerable attention recently in the context of twisted bilayer graphene~\cite{Khalaf2021a,Khalaf2021,Mai2022,Chatterjee2022} and related synthetic bilayer systems~\cite{Cian2020a}.
However, such experiments mainly use transport and spectroscopic measurements and therefore do not provide microscopic insights.
In contrast, quantum simulations allow for a better microscopic understanding, in particular of density and spin structures, and may provide direct coherent control over individual excitations, therefore allowing for a systematic study of their interactions.

Here we propose to realize Stoner ferromagnetism with ultracold fermions in optical lattices.
Cold atom experiments have already enabled extensive studies of the Fermi-Hubbard model~\cite{Joerdens2008,Esslinger2010,Tarruell2018,Bohrdt2021}, and implementations of artificial gauge fields in optical lattices exist~\cite{Aidelsburger2013,Miyake2013,Tai2017,Goldman2016,Aidelsburger2018}.
Combining both achievements allows to study ferromagnetism in the strongly doped Hubbard model and explore its relation to phenomena known to arise in high-$T_{\rm c}$ superconductors.
While textbook calculations predict Stoner ferromagnetism even without a magnetic field and analytical considerations for flat-band Hubbard models found ferromagnetic ground states in specific cases~\cite{Katsura2010}, we demonstrate that the tunability of synthetic magnetic fields in cold atoms~\cite{Tai2017} enables a systematic search for the onset of Stoner magnetism, the associated topological excitations and their interactions.

\begin{figure}[b]
	\centering
	\includegraphics{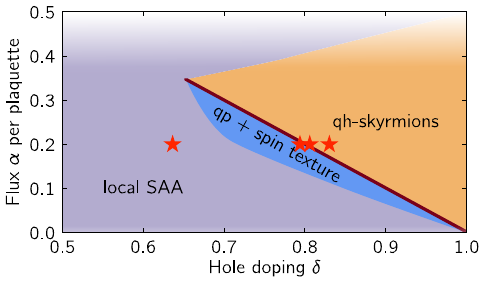}
	\caption{
		\label{fig:PhaseDiagram}
		Schematic phase diagram for the $S=0$ sector.
		We find quasi-hole (qh) skyrmion states, quasi-particle (qp) states carrying interesting spin textures, as well as states exhibiting local spin anti-alignment (SAA).
		While the qh-skyrmion state is a low-lying excitation of the QH ferromagnet, the other states constitute the global ground state of the model across all spin sectors.
		The stars indicate the points studied in Fig.~\ref{fig:Correlations}.
	}
\end{figure}

We study the strongly interacting Hofstadter-Fermi-Hubbard model, which provides an appropriate description of lattice fermions in a synthetic gauge field.
We perform density-matrix renormalization group (DMRG) simulations at variable flux per plaquette and hole doping.
In particular, we exploit the $\mathrm{SU}(2)$ spin-rotational symmetry of the model to study variational ground states in the fully spin polarized and the spin-singlet sectors.

We find that for magnetic filling $\nu\leq1$ the ground state of the Hofstadter-Fermi-Hubbard model is fully spin polarized for a wide range of parameters, realizing a lattice version of the QH~ferromagnet.
Furthermore, we observe spin textures reminiscent of skyrmions in the low-energy spin-singlet states, which for some parameters constitute the overall ground state.
Finally, for large magnetic flux per plaquette and lower doping, we predict the breakdown of Stoner ferromagnetism and observe ground states with local spin anti-alignment.
Our extracted phase diagram summarizes our findings in Fig.~\ref{fig:PhaseDiagram}.

\paragraph{Model.---}
We study spin-$\nicefrac{1}{2}$ fermions subject to a magnetic field on a two-dimensional square lattice of size $L_x \times L_y$.
Choosing the Landau gauge along the $y$-direction, the Hamiltonian reads
\begin{equation}
	\begin{aligned}
		\hat{\mathcal{H}} =
		&\sum_{x,y} \left[- t \sum_{\sigma}\left(\hat{c}^{\dagger}_{x+1, y, \sigma} \hat{c}^{\vphantom{\dagger}}_{x,y, \sigma} + \mathrm{e}^{2\pi i \alpha x} \hat{c}^{\dagger}_{x,y+1, \sigma} \hat{c}^{\vphantom{\dagger}}_{x,y, \sigma} \right.\right.\\
		&\phantom{\sum_{x,y} \left[- t \sum_{\sigma}\left(\right.\right.} \left.\left. \vphantom{\mathrm{e}^{2\pi i \alpha x} \hat{c}^{\dagger}_{x,y+1, \sigma}} + \mathrm{H.c.}\right) + \frac{U}{2} \hat{n}_{x,y, \uparrow} \hat{n}_{x,y, \downarrow}\right],
		\label{Eq:HBH-Hamiltonian}
	\end{aligned}
\end{equation}
where $\hat{c}_{x,y,\sigma}^{(\dagger)}$ is the annihilation (creation) operator for a spin-$\sigma$ fermion at site $(x,y)$ and $\hat{n}_{x,y,\sigma}$ is the corresponding number operator.
The first term of the Hamiltonian describes hopping on the lattice.
When hopping around a single plaquette, the fermions pick up a phase $2\pi \alpha$, corresponding to $\alpha$ flux quanta per plaquette.
The second term describes the Hubbard interaction of strength $U$ on doubly occupied sites.

We study the model on a cylinder by imposing open boundary conditions in $x$-direction and periodic boundary conditions in $y$-direction.
Furthermore, we choose strong repulsive Hubbard interactions, $\nicefrac{U}{t} = 8$, which includes Landau level mixing effects and is realistically achievable in state-of-the-art cold atom experiments.
However, we believe our results to extend to other values of the interaction strength exceeding the bandwidth.
For a system of $N$ particles we define the density $n = \nicefrac{N}{(L_x L_y)}$ and the hole concentration $\delta$ away from half-filling, such that $n=1-\delta$.
Furthermore, we define the magnetic filling factor $\nu = \nicefrac{N}{N_{\phi}}$,
where $N_{\phi} = \alpha \left(L_x-1\right) L_y$ is the total number of flux quanta.

In the limit of vanishing flux, $\alpha=0$, the model reduces to the 2D Hubbard model with its various quantum phases~\cite{Bohrdt2021,Arovas2022}.
Upon increasing the magnetic flux per plaquette, the single-particle Bloch bands split up into flat magnetic subbands resembling the Landau levels known from the continuum~\cite{Hofstadter1976}.
The formation of flat magnetic bands is expected to enhance interaction effects and leads to QH~physics, in particular the emergence of QH~ferromagnetism at filling factor $\nu=1$.
For small flux $\alpha$ and density $n$, lattice effects are negligible and we expect our results to connect to known continuum results~\cite{Palm2020}.
Increasing the flux per plaquette, lattice effects become more dominant and the continuum description breaks down.

To study lattice effects, we perform single-site DMRG simulations~\cite{White1992,Schollwoeck2011,Hubig2015,HubigSyTen}.
We exploit the $\mathrm{U}(1)$ symmetry associated with particle number conservation and the $\mathrm{SU}(2)$ spin-rotational symmetry of the model.
This allows us to calculate the variational ground state in the sectors of minimal ($S=0$) and maximal total spin ($S=S_{\rm max}=N/2$) on cylinders of size $L_x\times L_y = 31\times4$ and $33\times5$, and vary both the magnetic flux per plaquette and the density.

\begin{figure}[t]
	\centering
	\includegraphics{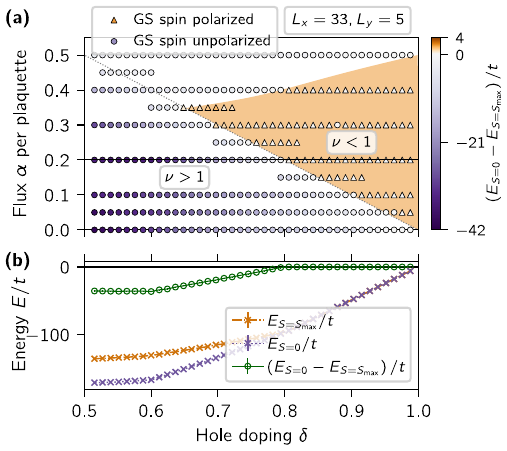}
	\caption{
		\label{fig:GSEnergies}
		(a) Energy difference between the lowest energy states found in DMRG for the $S=S_{\rm max}$ and the $S=0$ sectors as function of magnetic flux per plaquette $\alpha$ and doping level $\delta$.
		The gray dotted line indicates $\nu=1$.
		For $\alpha \lesssim 0.35$ and $\nu\leq1$ (shaded region) the ground state is spin polarized with an almost degenerate spin-singlet excited state (see also (b)).
		In contrast, for $\nu > 1$ the spin-singlet is energetically favored significantly.
		At large flux, $\alpha \gtrsim \alpha_c \approx 0.4$, the QH ferromagnetism breaks down and we find the ground state to be unpolarized even for $\nu<1$.
		(b) Ground state energies in both sectors at $\alpha=0.2$ (solid line in (a)).
		Data is given for a system of size $L_x\times L_y=33\times 5$.
	}
\end{figure}

\paragraph{QH Ferromagnetism at $\nu\lesssim1$.---}
First, we focus on systems with magnetic filling factor $\nu \approx 1$.
Close to the continuum limit, where the flux per plaquette $\alpha$ and the particle density $n=1-\delta$ are small, we expect the ground state to exhibit QH~ferromagnetism for $\nu=1$~\cite{Palm2020}.
This is confirmed by our DMRG simulations where we find the ground state to be spin polarized, see Fig.~\ref{fig:GSEnergies}.
This behavior extends significantly to larger flux $\alpha$ where lattice effects become increasingly relevant.
Additionally, in the spin polarized sector the system is incompressible at filling factor $\nu=1$ as expected for the corresponding QH~state, see Supplemental Material~\cite{supp}.
This clearly shows the emergence of QH~ferromagnetism in the Hofstadter-Fermi-Hubbard model.

Above some large critical flux $\alpha_c \approx 0.35$ per plaquette the QH~ferromagnetism at $\nu \lesssim 1$ breaks down and we find that the ground state is a spin-singlet.

\paragraph{Comparison with Trial States.---}
Before deepening our understanding of the DMRG results, we introduce two paradigmatic trial states describing the spin polarized and unpolarized sector, respectively.
Since the two states feature different short-range correlations, comparing their variational energies provides insights into the underlying magnetic order.

For the spin polarized case our trial state $\ket{\Psi^{\rm trial}_{S=S_{\rm max}}}$ is the exact eigenstate with energy $E^{\rm trial}_{S=S_{\rm max}}$ constituted by a Fermi sea of $N$ identical fermions.

For the spin-singlet sector we make a resonating valence bond (RVB) ansatz~\cite{Anderson1987}.
We start from a state consisting of $\nicefrac{N}{2}$ up- and down-spin fermions each forming separate Fermi seas, $\ket{\Psi^{\uparrow/\downarrow}}$.
To account for the strong Hubbard repulsion, we perform a Gutzwiller projection~\cite{Gutzwiller1963} to project out doubly occupied sites and obtain the trial state
\begin{equation}
	\begin{aligned}
		\ket{\Psi^{\rm trial}_{S=0}} &= \hat{\mathcal{P}}_{\rm G} \left(\ket{\Psi^{\uparrow}}\otimes\ket{\Psi^{\downarrow}}\right),\\
		\hat{\mathcal{P}}_{\rm G} &= \prod_{x,y} \left(1 - \hat{n}_{x,y,\uparrow}\hat{n}_{x,y,\downarrow}\right).
	\end{aligned}
\end{equation}
We determine the variational energy $E^{\rm trial}_{S=0}$ of this state using Metropolis Monte Carlo sampling of $N_{\rm MC}=1000$ snapshots.
\begin{figure}
	\centering
	\includegraphics{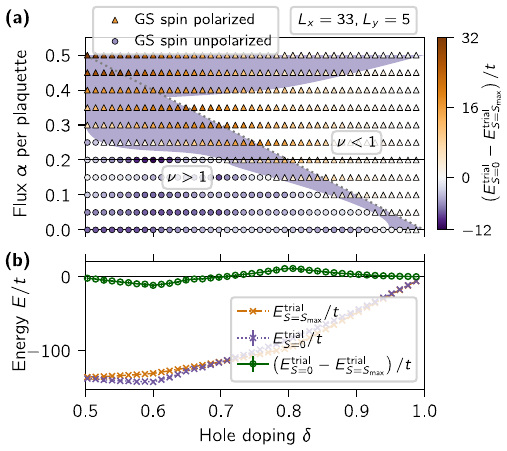}
	\caption{
		\label{fig:GSTrialStates}
		(a) Energy difference between the trial energies for the $S=S_{\rm max}$ and the $S=0$ sectors.
		The gray dotted line indicates $\nu = 1$ and the shaded area indicates the regime where the DMRG finds the ground state in the spin-singlet sector, while the trial states predict ferromagnetic order.
		(b) Trial state energies in both sectors at $\alpha=0.2$ (solid line in (a)).
		Data is given for a system of size $L_x\times L_y=33\times 5$.
	}
\end{figure}

As previously for the DMRG energies, we compare the trial energies in the different spin sectors in Fig.~\ref{fig:GSTrialStates}.
%
The trial states predict correctly the existence of QH ferromagnetism revealed earlier by DMRG.
For large $\alpha$ and some fillings $\nu \gtrsim 1$, however, we find a region (shading in Fig.~\ref{fig:GSTrialStates}) where QH ferromagnetism is predicted, but DMRG revealed a spin-singlet ground state.

The variational energy of the trial states is only sensitive to local correlations.
Hence the extended range of QH ferromagnetism predicted by the trial states provides a first indication that non-trivial spin textures may lead to the formation of spin-singlet ground states while retaining local ferromagnetism: this is a hallmark of skyrmion formation.

\paragraph{Skyrmions at $\nu \gtrless 1$.---}
We now turn to the low-energy spin-singlet states competing with the QH~ferromagnet.
In the regime we have just identified using the trial states, where local ferromagnetic correlations play an important role, we expect the spin-singlet states to exhibit interesting spin correlations evolving from short-range spin alignment into long-range anti-alignment.
A general overview of the resulting phase diagram for the $S=0$ sector is given in Fig.~\ref{fig:PhaseDiagram}.
To identify and understand the different phases, we consider the local density $n(x) = \sum_{y=1}^{L_y} \left\langle \hat{n}_{x,y}\right\rangle/L_y$ and the normalized spin-spin correlations
\begin{equation}
	C_{x_0}(x) = \frac{1}{L_y}\sum_{y=1}^{L_y}\left. \left\langle \hat{\vec{S}}_{x_0, y} \cdot \hat{\vec{S}}_{x,y} \right\rangle \right/ \left\langle \hat{n}_{x_0, y} \hat{n}_{x, y}\right\rangle
\end{equation}
relative to a fixed position $x_0=(L_x+1)/2$ in the bulk of the system.
On-site, we expect this spin-spin correlation function to be $C_{x_0}(x_0) = \nicefrac{3}{4}$ for spin-$\nicefrac{1}{2}$ fermions.

\begin{figure}[t]
	\centering
	\includegraphics{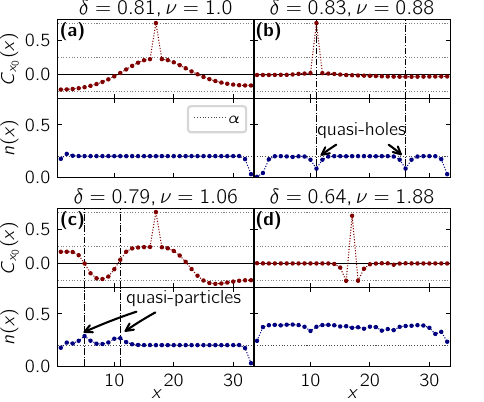}
	\caption{
		\label{fig:Correlations}
		Low-energy $S=0$ state at $\alpha=0.2$ with signatures of skyrmion states in (a-c).
		(a) $\nu = 1$: Spin-spin correlations resembling the characteristic behavior of skyrmions.
		(b) $\nu < 1$: Quasi-hole (qh) skyrmion state descending from the $\nu=1$ state.
		(c) $\nu > 1$: Quasi-particle (qp) skyrmion excitation of the $\nu=1$ state.
		(d) $\nu \gg 1$: State exhibiting local spin anti-alignment (SAA).
		Data is given for a system of size $L_x\times L_y = 33 \times 5$ and $x_0 = 17$ for the parameters indicated by stars in Fig.~\ref{fig:PhaseDiagram}.
	}
\end{figure}

At filling factor $\nu=1$, the low-energy spin-singlet excitation shows local ferromagnetic correlations, which at long distances continuously evolve into anti-aligned correlations, see Fig.~\ref{fig:Correlations}(a).
Similar to the continuum case~\cite{Lee1990,Sondhi1993}, this indicates the existence of skyrmion states in lattice systems.

Upon removing fermions from the spin-singlet $\nu=1$ state, the local density develops localized drops, see Fig.~\ref{fig:Correlations}(b).
We interpret these as quasi-holes of charge $q_{\rm qh}=-1$ derived from the $\nu=1$ state.
The charge of these quasi-holes is consistent with the prediction for skyrmions in the continuum QH~system~\cite{Sondhi1993}.
Furthermore, the spin-spin correlations around the quasi-holes are again consistent with localized skyrmion excitations.
Therefore, we interpret the low-lying excitations of the QH~ferromagnet at filling factor $\nu < 1$ as quasi-hole skyrmions.
In particular, we note that the number of quasi-hole skyrmions hosted by the system increases as the particle density is reduced.
We stress again that the overall ground state in this regime is spin polarized, while the spin-singlet states just discussed are low energy excitations.

In contrast, for filling factors $\nu > 1$, as additional fermions are added to the system, we always find the ground state to be in the $S=0$ sector independent of $\alpha$.
Furthermore, we find that the ground state hosts localized quasi-particles of charge $q_{\rm qp}=1$.
Similar to the quasi-hole case, we find a characteristic change of the spin-spin correlations around the quasi-particles as visualized in Fig.~\ref{fig:Correlations}(c).
While we can clearly identify these as quasi-hole skyrmions in the former case, the situation is less clear for the quasi-particles.
However, we find the quasi-particles to be accompanied by non-trivial spin-spin correlations when measured relative to a distant site in the bulk.
We interpret these quasi-particles as being dressed by non-trivial spin textures, which may be related to skyrmions or constitute a pre-cursor of stripe formation.

We note that the parameter regime where these exotic spin structures are observed is also part of the regime where our trial states predict the ground state to be ferromagnetically ordered globally.
In the ground states obtained using DMRG, this global order is replaced by local ferromagnetic correlations, while non-trivial spin textures lead to the formation of a global spin-singlet.

\paragraph{Non-skyrmion ground states.---}
So far, the states discussed were 
significantly influenced by the local spin alignment underlying the $\nu=1$ QH~ferromagnet.
Now, we turn to a first explorative analysis of the phase diagram in regimes where the influence of the QH~ferromagnetism essentially disappears.

Upon increasing the particle number such that $\nu \gg 1$, we find the ground state to be a spin-singlet characterized by local spin anti-alignment (SAA) with essentially uncorrelated spins on large length scales, see Fig.~\ref{fig:Correlations}(d).
This is in clear contrast to the skyrmion states which exhibit local ferromagnetism as one of their main features.
The local spin anti-alignment is reminiscent of free fermions exhibiting a Pauli exclusion hole~\cite{Koepsell2021}.

This state is the ground state for a broad range of parameters and might host other more complicated structures, some of which might be related to the myriads of phases known from the doped Hubbard model.
Furthermore, earlier studies~\cite{Tu2018} using renormalized mean-field theory and exact diagonalization also found phases with enlarged unit cells in this regime, which we can neither confirm nor completely exclude in our finite-size simulations.
Note that in this regime of large particle numbers the numerical simulations are particularly challenging and we believe the density to be not fully converged.

We now return to the regime $\nu \lesssim 1$. Upon increasing the flux per plaquette, we find a qualitatively change of the behavior.
For large flux per plaquette, $\alpha > \alpha_c \approx 0.35$, the QH~ferromagnetism breaks down and also the ground state at $\nu \lesssim 1$ is a spin-singlet.
Furthermore, for some values of the filling factor, this spin-singlet state does no longer exhibit neither skyrmionic correlations nor local SAA, but instead some oscillatory behavior of the spin correlations.

At the same time, in the very dilute limit, $\nu~\ll~1$, the ground state is again spin polarized even for large flux.
In this regime, spin textures reminiscent of the skyrmions discussed above become visible close to the transition region from spin polarized to spin-singlet ground states~\cite{supp}.

\paragraph{Conclusions and outlook.---}
We have found the quantum Hall ferromagnet to be the ground state of the Hofstadter-Fermi-Hubbard model at magnetic filling factor $\nu=1$.
Additionally, we have identified a skyrmion-like behavior of the low energy spin-singlet states.
In particular, the local density and the spin-spin correlations reveal quasi-hole skyrmions and quasi-particles dressed by spin textures.
For large flux per plaquette $\alpha \gtrsim 0.35$, we have observed a breakdown of QH ferromagnetism in favor of a spin-singlet ground state.
The microscopic nature of this state and its origin deserve a more detailed analysis and will be addressed in the future.

Our work paves the way for future investigations of many interacting skyrmions, in particular at very low fermion densities.
Furthermore, varying the Hubbard interaction strength $\nicefrac{U}{t}$ might give insight into the stability of the QH~ferromagnet against Landau level mixing or allow to explore doped chiral quantum spin liquids~\cite{Szasz2020}.
More broadly, connections to high-$T_{\rm c}$ superconductivity can be explored by going to the small-flux limit.
Extending the model to bilayer systems or finite temperatures will lead to further interesting questions for future research.
Furthermore, different lattice geometries exhibiting topological flat bands might allow for similar effects, even in the absence of a net magnetic field.

The model we studied can be realized with ultracold fermions in optical lattices, which provide simultaneous spin and charge resolution down to individual lattice sites.
In particular, specific spin sectors can be experimentally addressed using adiabatic preparation schemes~\cite{Palm2020}.
This type of system would also allow to add further-range dipolar interactions, which can be used to address the fractional~QH~regime where we expect similar skyrmion excitations to exist~\cite{Kamilla1996,Wojs2002,Doretto2005,Balram2015}.

\begin{acknowledgments}
\ 

We would like to thank  L.~Palm, P.~Preiss, M.~Rizzi, H.~Schl\"omer, and L.~Stenzel for fruitful discussions.
We acknowledge funding by the Deutsche Forschungsgemeinschaft (DFG, German Research Foundation) under Germany's Excellence Strategy -- EXC-2111 -- 390814868, and via Research Unit FOR 2414 under project number 277974659.
We acknowledge funding from the European Research Council (ERC) under the European Union’s Horizon 2020 research and innovation programm (Grant Agreement no 948141) -- ERC Starting Grant SimUcQuam.
AB acknowledges funding by the NSF through a grant for the Institute for Theoretical Atomic, Molecular, and Optical Physics at Harvard University and the Smithsonian Astrophysical Observatory.
\end{acknowledgments}

\bibliographystyle{apsrev4-1}

\begin{thebibliography}{49}%
	\makeatletter
	\providecommand \@ifxundefined [1]{%
		\@ifx{#1\undefined}
	}%
	\providecommand \@ifnum [1]{%
		\ifnum #1\expandafter \@firstoftwo
		\else \expandafter \@secondoftwo
		\fi
	}%
	\providecommand \@ifx [1]{%
		\ifx #1\expandafter \@firstoftwo
		\else \expandafter \@secondoftwo
		\fi
	}%
	\providecommand \natexlab [1]{#1}%
	\providecommand \enquote  [1]{``#1''}%
	\providecommand \bibnamefont  [1]{#1}%
	\providecommand \bibfnamefont [1]{#1}%
	\providecommand \citenamefont [1]{#1}%
	\providecommand \href@noop [0]{\@secondoftwo}%
	\providecommand \href [0]{\begingroup \@sanitize@url \@href}%
	\providecommand \@href[1]{\@@startlink{#1}\@@href}%
	\providecommand \@@href[1]{\endgroup#1\@@endlink}%
	\providecommand \@sanitize@url [0]{\catcode `\\12\catcode `\$12\catcode
		`\&12\catcode `\#12\catcode `\^12\catcode `\_12\catcode `\%12\relax}%
	\providecommand \@@startlink[1]{}%
	\providecommand \@@endlink[0]{}%
	\providecommand \url  [0]{\begingroup\@sanitize@url \@url }%
	\providecommand \@url [1]{\endgroup\@href {#1}{\urlprefix }}%
	\providecommand \urlprefix  [0]{URL }%
	\providecommand \Eprint [0]{\href }%
	\providecommand \doibase [0]{http://dx.doi.org/}%
	\providecommand \selectlanguage [0]{\@gobble}%
	\providecommand \bibinfo  [0]{\@secondoftwo}%
	\providecommand \bibfield  [0]{\@secondoftwo}%
	\providecommand \translation [1]{[#1]}%
	\providecommand \BibitemOpen [0]{}%
	\providecommand \bibitemStop [0]{}%
	\providecommand \bibitemNoStop [0]{.\EOS\space}%
	\providecommand \EOS [0]{\spacefactor3000\relax}%
	\providecommand \BibitemShut  [1]{\csname bibitem#1\endcsname}%
	\let\auto@bib@innerbib\@empty
	\bibitem [{\citenamefont {Yang}\ \emph {et~al.}(1994)\citenamefont {Yang},
		\citenamefont {Moon}, \citenamefont {Zheng}, \citenamefont {MacDonald},
		\citenamefont {Girvin}, \citenamefont {Yoshioka},\ and\ \citenamefont
		{Zhang}}]{Yang1994}%
	\BibitemOpen
	\bibfield  {author} {\bibinfo {author} {\bibfnamefont {K.}~\bibnamefont
			{Yang}}, \bibinfo {author} {\bibfnamefont {K.}~\bibnamefont {Moon}}, \bibinfo
		{author} {\bibfnamefont {L.}~\bibnamefont {Zheng}}, \bibinfo {author}
		{\bibfnamefont {A.~H.}\ \bibnamefont {MacDonald}}, \bibinfo {author}
		{\bibfnamefont {S.~M.}\ \bibnamefont {Girvin}}, \bibinfo {author}
		{\bibfnamefont {D.}~\bibnamefont {Yoshioka}}, \ and\ \bibinfo {author}
		{\bibfnamefont {S.-C.}\ \bibnamefont {Zhang}},\ }\href {\doibase
		10.1103/physrevlett.72.732} {\bibfield  {journal} {\bibinfo  {journal}
			{Physical Review Letters}\ }\textbf {\bibinfo {volume} {72}},\ \bibinfo
		{pages} {732} (\bibinfo {year} {1994})}\BibitemShut {NoStop}%
	\bibitem [{\citenamefont {Ezawa}\ and\ \citenamefont
		{Tsitsishvili}(2009)}]{Ezawa2009}%
	\BibitemOpen
	\bibfield  {author} {\bibinfo {author} {\bibfnamefont {Z.~F.}\ \bibnamefont
			{Ezawa}}\ and\ \bibinfo {author} {\bibfnamefont {G.}~\bibnamefont
			{Tsitsishvili}},\ }\href {\doibase 10.1088/0034-4885/72/8/086502} {\bibfield
		{journal} {\bibinfo  {journal} {Reports on Progress in Physics}\ }\textbf
		{\bibinfo {volume} {72}},\ \bibinfo {pages} {086502} (\bibinfo {year}
		{2009})}\BibitemShut {NoStop}%
	\bibitem [{\citenamefont {Rezayi}(1987)}]{Rezayi1987}%
	\BibitemOpen
	\bibfield  {author} {\bibinfo {author} {\bibfnamefont {E.~H.}\ \bibnamefont
			{Rezayi}},\ }\href {\doibase 10.1103/physrevb.36.5454} {\bibfield  {journal}
		{\bibinfo  {journal} {Physical Review B}\ }\textbf {\bibinfo {volume} {36}},\
		\bibinfo {pages} {5454} (\bibinfo {year} {1987})}\BibitemShut {NoStop}%
	\bibitem [{\citenamefont {Haldane}\ and\ \citenamefont
		{Rezayi}(1988)}]{Haldane1988}%
	\BibitemOpen
	\bibfield  {author} {\bibinfo {author} {\bibfnamefont {F.~D.~M.}\
			\bibnamefont {Haldane}}\ and\ \bibinfo {author} {\bibfnamefont {E.~H.}\
			\bibnamefont {Rezayi}},\ }\href {\doibase 10.1103/physrevlett.60.956}
	{\bibfield  {journal} {\bibinfo  {journal} {Physical Review Letters}\
		}\textbf {\bibinfo {volume} {60}},\ \bibinfo {pages} {956} (\bibinfo {year}
		{1988})}\BibitemShut {NoStop}%
	\bibitem [{\citenamefont {Lee}\ and\ \citenamefont {Kane}(1990)}]{Lee1990}%
	\BibitemOpen
	\bibfield  {author} {\bibinfo {author} {\bibfnamefont {D.-H.}\ \bibnamefont
			{Lee}}\ and\ \bibinfo {author} {\bibfnamefont {C.~L.}\ \bibnamefont {Kane}},\
	}\href {\doibase 10.1103/physrevlett.64.1313} {\bibfield  {journal} {\bibinfo
			{journal} {Physical Review Letters}\ }\textbf {\bibinfo {volume} {64}},\
		\bibinfo {pages} {1313} (\bibinfo {year} {1990})}\BibitemShut {NoStop}%
	\bibitem [{\citenamefont {Rezayi}(1991)}]{Rezayi1991}%
	\BibitemOpen
	\bibfield  {author} {\bibinfo {author} {\bibfnamefont {E.~H.}\ \bibnamefont
			{Rezayi}},\ }\href {\doibase 10.1103/physrevb.43.5944} {\bibfield  {journal}
		{\bibinfo  {journal} {Physical Review B}\ }\textbf {\bibinfo {volume} {43}},\
		\bibinfo {pages} {5944} (\bibinfo {year} {1991})}\BibitemShut {NoStop}%
	\bibitem [{\citenamefont {Sondhi}\ \emph {et~al.}(1993)\citenamefont {Sondhi},
		\citenamefont {Karlhede}, \citenamefont {Kivelson},\ and\ \citenamefont
		{Rezayi}}]{Sondhi1993}%
	\BibitemOpen
	\bibfield  {author} {\bibinfo {author} {\bibfnamefont {S.~L.}\ \bibnamefont
			{Sondhi}}, \bibinfo {author} {\bibfnamefont {A.}~\bibnamefont {Karlhede}},
		\bibinfo {author} {\bibfnamefont {S.~A.}\ \bibnamefont {Kivelson}}, \ and\
		\bibinfo {author} {\bibfnamefont {E.~H.}\ \bibnamefont {Rezayi}},\ }\href
	{\doibase 10.1103/physrevb.47.16419} {\bibfield  {journal} {\bibinfo
			{journal} {Physical Review B}\ }\textbf {\bibinfo {volume} {47}},\ \bibinfo
		{pages} {16419} (\bibinfo {year} {1993})}\BibitemShut {NoStop}%
	\bibitem [{\citenamefont {Fertig}\ \emph {et~al.}(1994)\citenamefont {Fertig},
		\citenamefont {Brey}, \citenamefont {C{\^{o}}t{\'{e}}},\ and\ \citenamefont
		{MacDonald}}]{Fertig1994}%
	\BibitemOpen
	\bibfield  {author} {\bibinfo {author} {\bibfnamefont {H.~A.}\ \bibnamefont
			{Fertig}}, \bibinfo {author} {\bibfnamefont {L.}~\bibnamefont {Brey}},
		\bibinfo {author} {\bibfnamefont {R.}~\bibnamefont {C{\^{o}}t{\'{e}}}}, \
		and\ \bibinfo {author} {\bibfnamefont {A.~H.}\ \bibnamefont {MacDonald}},\
	}\href {\doibase 10.1103/physrevb.50.11018} {\bibfield  {journal} {\bibinfo
			{journal} {Physical Review B}\ }\textbf {\bibinfo {volume} {50}},\ \bibinfo
		{pages} {11018} (\bibinfo {year} {1994})}\BibitemShut {NoStop}%
	\bibitem [{\citenamefont {Moon}\ \emph {et~al.}(1995)\citenamefont {Moon},
		\citenamefont {Mori}, \citenamefont {Yang}, \citenamefont {Girvin},
		\citenamefont {MacDonald}, \citenamefont {Zheng}, \citenamefont {Yoshioka},\
		and\ \citenamefont {Zhang}}]{Moon1995}%
	\BibitemOpen
	\bibfield  {author} {\bibinfo {author} {\bibfnamefont {K.}~\bibnamefont
			{Moon}}, \bibinfo {author} {\bibfnamefont {H.}~\bibnamefont {Mori}}, \bibinfo
		{author} {\bibfnamefont {K.}~\bibnamefont {Yang}}, \bibinfo {author}
		{\bibfnamefont {S.~M.}\ \bibnamefont {Girvin}}, \bibinfo {author}
		{\bibfnamefont {A.~H.}\ \bibnamefont {MacDonald}}, \bibinfo {author}
		{\bibfnamefont {L.}~\bibnamefont {Zheng}}, \bibinfo {author} {\bibfnamefont
			{D.}~\bibnamefont {Yoshioka}}, \ and\ \bibinfo {author} {\bibfnamefont
			{S.-C.}\ \bibnamefont {Zhang}},\ }\href {\doibase 10.1103/physrevb.51.5138}
	{\bibfield  {journal} {\bibinfo  {journal} {Physical Review B}\ }\textbf
		{\bibinfo {volume} {51}},\ \bibinfo {pages} {5138} (\bibinfo {year}
		{1995})}\BibitemShut {NoStop}%
	\bibitem [{\citenamefont {MacDonald}\ \emph {et~al.}(1996)\citenamefont
		{MacDonald}, \citenamefont {Fertig},\ and\ \citenamefont
		{Brey}}]{MacDonald1996}%
	\BibitemOpen
	\bibfield  {author} {\bibinfo {author} {\bibfnamefont {A.~H.}\ \bibnamefont
			{MacDonald}}, \bibinfo {author} {\bibfnamefont {H.~A.}\ \bibnamefont
			{Fertig}}, \ and\ \bibinfo {author} {\bibfnamefont {L.}~\bibnamefont
			{Brey}},\ }\href {\doibase 10.1103/physrevlett.76.2153} {\bibfield  {journal}
		{\bibinfo  {journal} {Physical Review Letters}\ }\textbf {\bibinfo {volume}
			{76}},\ \bibinfo {pages} {2153} (\bibinfo {year} {1996})}\BibitemShut
	{NoStop}%
	\bibitem [{\citenamefont {Barrett}\ \emph {et~al.}(1995)\citenamefont
		{Barrett}, \citenamefont {Dabbagh}, \citenamefont {Pfeiffer}, \citenamefont
		{West},\ and\ \citenamefont {Tycko}}]{Barrett1995}%
	\BibitemOpen
	\bibfield  {author} {\bibinfo {author} {\bibfnamefont {S.~E.}\ \bibnamefont
			{Barrett}}, \bibinfo {author} {\bibfnamefont {G.}~\bibnamefont {Dabbagh}},
		\bibinfo {author} {\bibfnamefont {L.~N.}\ \bibnamefont {Pfeiffer}}, \bibinfo
		{author} {\bibfnamefont {K.~W.}\ \bibnamefont {West}}, \ and\ \bibinfo
		{author} {\bibfnamefont {R.}~\bibnamefont {Tycko}},\ }\href {\doibase
		10.1103/physrevlett.74.5112} {\bibfield  {journal} {\bibinfo  {journal}
			{Physical Review Letters}\ }\textbf {\bibinfo {volume} {74}},\ \bibinfo
		{pages} {5112} (\bibinfo {year} {1995})}\BibitemShut {NoStop}%
	\bibitem [{\citenamefont {Schmeller}\ \emph {et~al.}(1995)\citenamefont
		{Schmeller}, \citenamefont {Eisenstein}, \citenamefont {Pfeiffer},\ and\
		\citenamefont {West}}]{Schmeller1995}%
	\BibitemOpen
	\bibfield  {author} {\bibinfo {author} {\bibfnamefont {A.}~\bibnamefont
			{Schmeller}}, \bibinfo {author} {\bibfnamefont {J.~P.}\ \bibnamefont
			{Eisenstein}}, \bibinfo {author} {\bibfnamefont {L.~N.}\ \bibnamefont
			{Pfeiffer}}, \ and\ \bibinfo {author} {\bibfnamefont {K.~W.}\ \bibnamefont
			{West}},\ }\href {\doibase 10.1103/physrevlett.75.4290} {\bibfield  {journal}
		{\bibinfo  {journal} {Physical Review Letters}\ }\textbf {\bibinfo {volume}
			{75}},\ \bibinfo {pages} {4290} (\bibinfo {year} {1995})}\BibitemShut
	{NoStop}%
	\bibitem [{\citenamefont {Aifer}\ \emph {et~al.}(1996)\citenamefont {Aifer},
		\citenamefont {Goldberg},\ and\ \citenamefont {Broido}}]{Aifer1996}%
	\BibitemOpen
	\bibfield  {author} {\bibinfo {author} {\bibfnamefont {E.~H.}\ \bibnamefont
			{Aifer}}, \bibinfo {author} {\bibfnamefont {B.~B.}\ \bibnamefont {Goldberg}},
		\ and\ \bibinfo {author} {\bibfnamefont {D.~A.}\ \bibnamefont {Broido}},\
	}\href {\doibase 10.1103/physrevlett.76.680} {\bibfield  {journal} {\bibinfo
			{journal} {Physical Review Letters}\ }\textbf {\bibinfo {volume} {76}},\
		\bibinfo {pages} {680} (\bibinfo {year} {1996})}\BibitemShut {NoStop}%
	\bibitem [{\citenamefont {Manfra}\ \emph {et~al.}(1997)\citenamefont {Manfra},
		\citenamefont {Goldberg}, \citenamefont {Pfeiffer},\ and\ \citenamefont
		{West}}]{Manfra1997}%
	\BibitemOpen
	\bibfield  {author} {\bibinfo {author} {\bibfnamefont {M.~J.}\ \bibnamefont
			{Manfra}}, \bibinfo {author} {\bibfnamefont {B.~B.}\ \bibnamefont
			{Goldberg}}, \bibinfo {author} {\bibfnamefont {L.}~\bibnamefont {Pfeiffer}},
		\ and\ \bibinfo {author} {\bibfnamefont {K.}~\bibnamefont {West}},\ }\href
	{\doibase 10.1016/s1386-9477(97)00006-4} {\bibfield  {journal} {\bibinfo
			{journal} {Physica E: Low-dimensional Systems and Nanostructures}\ }\textbf
		{\bibinfo {volume} {1}},\ \bibinfo {pages} {28} (\bibinfo {year}
		{1997})}\BibitemShut {NoStop}%
	\bibitem [{\citenamefont {Townsley}\ \emph {et~al.}(2005)\citenamefont
		{Townsley}, \citenamefont {Chughtai}, \citenamefont {Nicholas},\ and\
		\citenamefont {Henini}}]{Townsley2005}%
	\BibitemOpen
	\bibfield  {author} {\bibinfo {author} {\bibfnamefont {C.~M.}\ \bibnamefont
			{Townsley}}, \bibinfo {author} {\bibfnamefont {R.}~\bibnamefont {Chughtai}},
		\bibinfo {author} {\bibfnamefont {R.~J.}\ \bibnamefont {Nicholas}}, \ and\
		\bibinfo {author} {\bibfnamefont {M.}~\bibnamefont {Henini}},\ }\href
	{\doibase 10.1103/physrevb.71.073303} {\bibfield  {journal} {\bibinfo
			{journal} {Physical Review B}\ }\textbf {\bibinfo {volume} {71}},\ \bibinfo
		{pages} {073303} (\bibinfo {year} {2005})}\BibitemShut {NoStop}%
	\bibitem [{\citenamefont {Bryja}\ \emph {et~al.}(2006)\citenamefont {Bryja},
		\citenamefont {W{\'{o}}js},\ and\ \citenamefont {Potemski}}]{Bryja2006}%
	\BibitemOpen
	\bibfield  {author} {\bibinfo {author} {\bibfnamefont {L.}~\bibnamefont
			{Bryja}}, \bibinfo {author} {\bibfnamefont {A.}~\bibnamefont {W{\'{o}}js}}, \
		and\ \bibinfo {author} {\bibfnamefont {M.}~\bibnamefont {Potemski}},\ }\href
	{\doibase 10.1103/physrevb.73.241302} {\bibfield  {journal} {\bibinfo
			{journal} {Physical Review B}\ }\textbf {\bibinfo {volume} {73}},\ \bibinfo
		{pages} {241302} (\bibinfo {year} {2006})}\BibitemShut {NoStop}%
	\bibitem [{\citenamefont {Lupatini}\ \emph {et~al.}(2020)\citenamefont
		{Lupatini}, \citenamefont {Knüppel}, \citenamefont {Faelt}, \citenamefont
		{Winkler}, \citenamefont {Shayegan}, \citenamefont {Imamoglu},\ and\
		\citenamefont {Wegscheider}}]{Lupatini2020}%
	\BibitemOpen
	\bibfield  {author} {\bibinfo {author} {\bibfnamefont {M.}~\bibnamefont
			{Lupatini}}, \bibinfo {author} {\bibfnamefont {P.}~\bibnamefont {Knüppel}},
		\bibinfo {author} {\bibfnamefont {S.}~\bibnamefont {Faelt}}, \bibinfo
		{author} {\bibfnamefont {R.}~\bibnamefont {Winkler}}, \bibinfo {author}
		{\bibfnamefont {M.}~\bibnamefont {Shayegan}}, \bibinfo {author}
		{\bibfnamefont {A.}~\bibnamefont {Imamoglu}}, \ and\ \bibinfo {author}
		{\bibfnamefont {W.}~\bibnamefont {Wegscheider}},\ }\href {\doibase
		10.1103/physrevlett.125.067404} {\bibfield  {journal} {\bibinfo  {journal}
			{Physical Review Letters}\ }\textbf {\bibinfo {volume} {125}},\ \bibinfo
		{pages} {067404} (\bibinfo {year} {2020})}\BibitemShut {NoStop}%
	\bibitem [{\citenamefont {Khalaf}\ \emph {et~al.}(2021)\citenamefont {Khalaf},
		\citenamefont {Chatterjee}, \citenamefont {Bultinck}, \citenamefont
		{Zaletel},\ and\ \citenamefont {Vishwanath}}]{Khalaf2021a}%
	\BibitemOpen
	\bibfield  {author} {\bibinfo {author} {\bibfnamefont {E.}~\bibnamefont
			{Khalaf}}, \bibinfo {author} {\bibfnamefont {S.}~\bibnamefont {Chatterjee}},
		\bibinfo {author} {\bibfnamefont {N.}~\bibnamefont {Bultinck}}, \bibinfo
		{author} {\bibfnamefont {M.~P.}\ \bibnamefont {Zaletel}}, \ and\ \bibinfo
		{author} {\bibfnamefont {A.}~\bibnamefont {Vishwanath}},\ }\href {\doibase
		10.1126/sciadv.abf5299} {\bibfield  {journal} {\bibinfo  {journal} {Science
				Advances}\ }\textbf {\bibinfo {volume} {7}} (\bibinfo {year} {2021}),\
		10.1126/sciadv.abf5299}\BibitemShut {NoStop}%
	\bibitem [{\citenamefont {Khalaf}\ and\ \citenamefont
		{Vishwanath}(2021)}]{Khalaf2021}%
	\BibitemOpen
	\bibfield  {author} {\bibinfo {author} {\bibfnamefont {E.}~\bibnamefont
			{Khalaf}}\ and\ \bibinfo {author} {\bibfnamefont {A.}~\bibnamefont
			{Vishwanath}},\ }\href@noop {} {\  (\bibinfo {year} {2021})},\ \Eprint
	{http://arxiv.org/abs/2112.06935} {arXiv:2112.06935} \BibitemShut {NoStop}%
	\bibitem [{\citenamefont {Mai}\ \emph {et~al.}(2022)\citenamefont {Mai},
		\citenamefont {Huang}, \citenamefont {Yu}, \citenamefont {Feldman},\ and\
		\citenamefont {Phillips}}]{Mai2022}%
	\BibitemOpen
	\bibfield  {author} {\bibinfo {author} {\bibfnamefont {P.}~\bibnamefont
			{Mai}}, \bibinfo {author} {\bibfnamefont {E.~W.}\ \bibnamefont {Huang}},
		\bibinfo {author} {\bibfnamefont {J.}~\bibnamefont {Yu}}, \bibinfo {author}
		{\bibfnamefont {B.~E.}\ \bibnamefont {Feldman}}, \ and\ \bibinfo {author}
		{\bibfnamefont {P.~W.}\ \bibnamefont {Phillips}},\ }\href@noop {} {\
		(\bibinfo {year} {2022})},\ \Eprint {http://arxiv.org/abs/2205.08545}
	{arXiv:2205.08545} \BibitemShut {NoStop}%
	\bibitem [{\citenamefont {Chatterjee}\ \emph {et~al.}(2022)\citenamefont
		{Chatterjee}, \citenamefont {Ippoliti},\ and\ \citenamefont
		{Zaletel}}]{Chatterjee2022}%
	\BibitemOpen
	\bibfield  {author} {\bibinfo {author} {\bibfnamefont {S.}~\bibnamefont
			{Chatterjee}}, \bibinfo {author} {\bibfnamefont {M.}~\bibnamefont
			{Ippoliti}}, \ and\ \bibinfo {author} {\bibfnamefont {M.~P.}\ \bibnamefont
			{Zaletel}},\ }\href {\doibase 10.1103/PhysRevB.106.035421} {\bibfield
		{journal} {\bibinfo  {journal} {Physical Review B}\ }\textbf {\bibinfo
			{volume} {106}},\ \bibinfo {pages} {035421} (\bibinfo {year}
		{2022})}\BibitemShut {NoStop}%
	\bibitem [{\citenamefont {Cian}\ \emph {et~al.}(2020)\citenamefont {Cian},
		\citenamefont {Grass}, \citenamefont {Vaezi}, \citenamefont {Liu},\ and\
		\citenamefont {Hafezi}}]{Cian2020a}%
	\BibitemOpen
	\bibfield  {author} {\bibinfo {author} {\bibfnamefont {Z.-P.}\ \bibnamefont
			{Cian}}, \bibinfo {author} {\bibfnamefont {T.}~\bibnamefont {Grass}},
		\bibinfo {author} {\bibfnamefont {A.}~\bibnamefont {Vaezi}}, \bibinfo
		{author} {\bibfnamefont {Z.}~\bibnamefont {Liu}}, \ and\ \bibinfo {author}
		{\bibfnamefont {M.}~\bibnamefont {Hafezi}},\ }\href {\doibase
		10.1103/physrevb.102.085430} {\bibfield  {journal} {\bibinfo  {journal}
			{Physical Review B}\ }\textbf {\bibinfo {volume} {102}},\ \bibinfo {pages}
		{085430} (\bibinfo {year} {2020})}\BibitemShut {NoStop}%
	\bibitem [{\citenamefont {Jördens}\ \emph {et~al.}(2008)\citenamefont
		{Jördens}, \citenamefont {Strohmaier}, \citenamefont {Günter},
		\citenamefont {Moritz},\ and\ \citenamefont {Esslinger}}]{Joerdens2008}%
	\BibitemOpen
	\bibfield  {author} {\bibinfo {author} {\bibfnamefont {R.}~\bibnamefont
			{Jördens}}, \bibinfo {author} {\bibfnamefont {N.}~\bibnamefont
			{Strohmaier}}, \bibinfo {author} {\bibfnamefont {K.}~\bibnamefont {Günter}},
		\bibinfo {author} {\bibfnamefont {H.}~\bibnamefont {Moritz}}, \ and\ \bibinfo
		{author} {\bibfnamefont {T.}~\bibnamefont {Esslinger}},\ }\href {\doibase
		10.1038/nature07244} {\bibfield  {journal} {\bibinfo  {journal} {Nature}\
		}\textbf {\bibinfo {volume} {455}},\ \bibinfo {pages} {204} (\bibinfo {year}
		{2008})}\BibitemShut {NoStop}%
	\bibitem [{\citenamefont {Esslinger}(2010)}]{Esslinger2010}%
	\BibitemOpen
	\bibfield  {author} {\bibinfo {author} {\bibfnamefont {T.}~\bibnamefont
			{Esslinger}},\ }\href {\doibase 10.1146/annurev-conmatphys-070909-104059}
	{\bibfield  {journal} {\bibinfo  {journal} {Annual Review of Condensed Matter
				Physics}\ }\textbf {\bibinfo {volume} {1}},\ \bibinfo {pages} {129} (\bibinfo
		{year} {2010})}\BibitemShut {NoStop}%
	\bibitem [{\citenamefont {Tarruell}\ and\ \citenamefont
		{Sanchez-Palencia}(2018)}]{Tarruell2018}%
	\BibitemOpen
	\bibfield  {author} {\bibinfo {author} {\bibfnamefont {L.}~\bibnamefont
			{Tarruell}}\ and\ \bibinfo {author} {\bibfnamefont {L.}~\bibnamefont
			{Sanchez-Palencia}},\ }\href {\doibase 10.1016/j.crhy.2018.10.013} {\bibfield
		{journal} {\bibinfo  {journal} {Comptes Rendus Physique}\ }\textbf {\bibinfo
			{volume} {19}},\ \bibinfo {pages} {365} (\bibinfo {year} {2018})}\BibitemShut
	{NoStop}%
	\bibitem [{\citenamefont {Bohrdt}\ \emph {et~al.}(2021)\citenamefont {Bohrdt},
		\citenamefont {Homeier}, \citenamefont {Reinmoser}, \citenamefont {Demler},\
		and\ \citenamefont {Grusdt}}]{Bohrdt2021}%
	\BibitemOpen
	\bibfield  {author} {\bibinfo {author} {\bibfnamefont {A.}~\bibnamefont
			{Bohrdt}}, \bibinfo {author} {\bibfnamefont {L.}~\bibnamefont {Homeier}},
		\bibinfo {author} {\bibfnamefont {C.}~\bibnamefont {Reinmoser}}, \bibinfo
		{author} {\bibfnamefont {E.}~\bibnamefont {Demler}}, \ and\ \bibinfo {author}
		{\bibfnamefont {F.}~\bibnamefont {Grusdt}},\ }\href {\doibase
		10.1016/j.aop.2021.168651} {\bibfield  {journal} {\bibinfo  {journal} {Annals
				of Physics}\ }\textbf {\bibinfo {volume} {435}},\ \bibinfo {pages} {168651}
		(\bibinfo {year} {2021})}\BibitemShut {NoStop}%
	\bibitem [{\citenamefont {Aidelsburger}\ \emph {et~al.}(2013)\citenamefont
		{Aidelsburger}, \citenamefont {Atala}, \citenamefont {Lohse}, \citenamefont
		{Barreiro}, \citenamefont {Paredes},\ and\ \citenamefont
		{Bloch}}]{Aidelsburger2013}%
	\BibitemOpen
	\bibfield  {author} {\bibinfo {author} {\bibfnamefont {M.}~\bibnamefont
			{Aidelsburger}}, \bibinfo {author} {\bibfnamefont {M.}~\bibnamefont {Atala}},
		\bibinfo {author} {\bibfnamefont {M.}~\bibnamefont {Lohse}}, \bibinfo
		{author} {\bibfnamefont {J.~T.}\ \bibnamefont {Barreiro}}, \bibinfo {author}
		{\bibfnamefont {B.}~\bibnamefont {Paredes}}, \ and\ \bibinfo {author}
		{\bibfnamefont {I.}~\bibnamefont {Bloch}},\ }\href {\doibase
		10.1103/physrevlett.111.185301} {\bibfield  {journal} {\bibinfo  {journal}
			{Physical Review Letters}\ }\textbf {\bibinfo {volume} {111}},\ \bibinfo
		{pages} {185301} (\bibinfo {year} {2013})}\BibitemShut {NoStop}%
	\bibitem [{\citenamefont {Miyake}\ \emph {et~al.}(2013)\citenamefont {Miyake},
		\citenamefont {Siviloglou}, \citenamefont {Kennedy}, \citenamefont {Burton},\
		and\ \citenamefont {Ketterle}}]{Miyake2013}%
	\BibitemOpen
	\bibfield  {author} {\bibinfo {author} {\bibfnamefont {H.}~\bibnamefont
			{Miyake}}, \bibinfo {author} {\bibfnamefont {G.~A.}\ \bibnamefont
			{Siviloglou}}, \bibinfo {author} {\bibfnamefont {C.~J.}\ \bibnamefont
			{Kennedy}}, \bibinfo {author} {\bibfnamefont {W.~C.}\ \bibnamefont {Burton}},
		\ and\ \bibinfo {author} {\bibfnamefont {W.}~\bibnamefont {Ketterle}},\
	}\href {\doibase 10.1103/physrevlett.111.185302} {\bibfield  {journal}
		{\bibinfo  {journal} {Physical Review Letters}\ }\textbf {\bibinfo {volume}
			{111}},\ \bibinfo {pages} {185302} (\bibinfo {year} {2013})}\BibitemShut
	{NoStop}%
	\bibitem [{\citenamefont {Tai}\ \emph {et~al.}(2017)\citenamefont {Tai},
		\citenamefont {Lukin}, \citenamefont {Rispoli}, \citenamefont {Schittko},
		\citenamefont {Menke}, \citenamefont {Borgnia}, \citenamefont {Preiss},
		\citenamefont {Grusdt}, \citenamefont {Kaufman},\ and\ \citenamefont
		{Greiner}}]{Tai2017}%
	\BibitemOpen
	\bibfield  {author} {\bibinfo {author} {\bibfnamefont {M.~E.}\ \bibnamefont
			{Tai}}, \bibinfo {author} {\bibfnamefont {A.}~\bibnamefont {Lukin}}, \bibinfo
		{author} {\bibfnamefont {M.}~\bibnamefont {Rispoli}}, \bibinfo {author}
		{\bibfnamefont {R.}~\bibnamefont {Schittko}}, \bibinfo {author}
		{\bibfnamefont {T.}~\bibnamefont {Menke}}, \bibinfo {author} {\bibfnamefont
			{D.}~\bibnamefont {Borgnia}}, \bibinfo {author} {\bibfnamefont {P.~M.}\
			\bibnamefont {Preiss}}, \bibinfo {author} {\bibfnamefont {F.}~\bibnamefont
			{Grusdt}}, \bibinfo {author} {\bibfnamefont {A.~M.}\ \bibnamefont {Kaufman}},
		\ and\ \bibinfo {author} {\bibfnamefont {M.}~\bibnamefont {Greiner}},\ }\href
	{\doibase 10.1038/nature22811} {\bibfield  {journal} {\bibinfo  {journal}
			{Nature}\ }\textbf {\bibinfo {volume} {546}},\ \bibinfo {pages} {519}
		(\bibinfo {year} {2017})}\BibitemShut {NoStop}%
	\bibitem [{\citenamefont {Goldman}\ \emph {et~al.}(2016)\citenamefont
		{Goldman}, \citenamefont {Budich},\ and\ \citenamefont
		{Zoller}}]{Goldman2016}%
	\BibitemOpen
	\bibfield  {author} {\bibinfo {author} {\bibfnamefont {N.}~\bibnamefont
			{Goldman}}, \bibinfo {author} {\bibfnamefont {J.~C.}\ \bibnamefont {Budich}},
		\ and\ \bibinfo {author} {\bibfnamefont {P.}~\bibnamefont {Zoller}},\ }\href
	{\doibase 10.1038/nphys3803} {\bibfield  {journal} {\bibinfo  {journal}
			{Nature Physics}\ }\textbf {\bibinfo {volume} {12}},\ \bibinfo {pages} {639}
		(\bibinfo {year} {2016})}\BibitemShut {NoStop}%
	\bibitem [{\citenamefont {Aidelsburger}\ \emph {et~al.}(2018)\citenamefont
		{Aidelsburger}, \citenamefont {Nascimbene},\ and\ \citenamefont
		{Goldman}}]{Aidelsburger2018}%
	\BibitemOpen
	\bibfield  {author} {\bibinfo {author} {\bibfnamefont {M.}~\bibnamefont
			{Aidelsburger}}, \bibinfo {author} {\bibfnamefont {S.}~\bibnamefont
			{Nascimbene}}, \ and\ \bibinfo {author} {\bibfnamefont {N.}~\bibnamefont
			{Goldman}},\ }\href {\doibase 10.1016/j.crhy.2018.03.002} {\bibfield
		{journal} {\bibinfo  {journal} {Comptes Rendus Physique}\ }\textbf {\bibinfo
			{volume} {19}},\ \bibinfo {pages} {394} (\bibinfo {year} {2018})}\BibitemShut
	{NoStop}%
	\bibitem [{\citenamefont {Katsura}\ \emph {et~al.}(2010)\citenamefont
		{Katsura}, \citenamefont {Maruyama}, \citenamefont {Tanaka},\ and\
		\citenamefont {Tasaki}}]{Katsura2010}%
	\BibitemOpen
	\bibfield  {author} {\bibinfo {author} {\bibfnamefont {H.}~\bibnamefont
			{Katsura}}, \bibinfo {author} {\bibfnamefont {I.}~\bibnamefont {Maruyama}},
		\bibinfo {author} {\bibfnamefont {A.}~\bibnamefont {Tanaka}}, \ and\ \bibinfo
		{author} {\bibfnamefont {H.}~\bibnamefont {Tasaki}},\ }\href {\doibase
		10.1209/0295-5075/91/57007} {\bibfield  {journal} {\bibinfo  {journal}
			{Europhysics Letters}\ }\textbf {\bibinfo {volume} {91}},\ \bibinfo {pages}
		{57007} (\bibinfo {year} {2010})}\BibitemShut {NoStop}%
	\bibitem [{\citenamefont {Arovas}\ \emph {et~al.}(2022)\citenamefont {Arovas},
		\citenamefont {Berg}, \citenamefont {Kivelson},\ and\ \citenamefont
		{Raghu}}]{Arovas2022}%
	\BibitemOpen
	\bibfield  {author} {\bibinfo {author} {\bibfnamefont {D.~P.}\ \bibnamefont
			{Arovas}}, \bibinfo {author} {\bibfnamefont {E.}~\bibnamefont {Berg}},
		\bibinfo {author} {\bibfnamefont {S.}~\bibnamefont {Kivelson}}, \ and\
		\bibinfo {author} {\bibfnamefont {S.}~\bibnamefont {Raghu}},\ }\href
	{\doibase 10.1146/annurev-conmatphys-031620-102024} {\bibfield  {journal}
		{\bibinfo  {journal} {Annual Review of Condensed Matter Physics}\ }\textbf
		{\bibinfo {volume} {13}},\ \bibinfo {pages} {239} (\bibinfo {year}
		{2022})}\BibitemShut {NoStop}%
	\bibitem [{\citenamefont {Hofstadter}(1976)}]{Hofstadter1976}%
	\BibitemOpen
	\bibfield  {author} {\bibinfo {author} {\bibfnamefont {D.~R.}\ \bibnamefont
			{Hofstadter}},\ }\href {\doibase 10.1103/physrevb.14.2239} {\bibfield
		{journal} {\bibinfo  {journal} {Physical Review B}\ }\textbf {\bibinfo
			{volume} {14}},\ \bibinfo {pages} {2239} (\bibinfo {year}
		{1976})}\BibitemShut {NoStop}%
	\bibitem [{\citenamefont {Palm}\ \emph {et~al.}(2020)\citenamefont {Palm},
		\citenamefont {Grusdt},\ and\ \citenamefont {Preiss}}]{Palm2020}%
	\BibitemOpen
	\bibfield  {author} {\bibinfo {author} {\bibfnamefont {L.}~\bibnamefont
			{Palm}}, \bibinfo {author} {\bibfnamefont {F.}~\bibnamefont {Grusdt}}, \ and\
		\bibinfo {author} {\bibfnamefont {P.~M.}\ \bibnamefont {Preiss}},\ }\href
	{\doibase 10.1088/1367-2630/aba30e} {\bibfield  {journal} {\bibinfo
			{journal} {New Journal of Physics}\ }\textbf {\bibinfo {volume} {22}},\
		\bibinfo {pages} {083037} (\bibinfo {year} {2020})}\BibitemShut {NoStop}%
	\bibitem [{\citenamefont {White}(1992)}]{White1992}%
	\BibitemOpen
	\bibfield  {author} {\bibinfo {author} {\bibfnamefont {S.~R.}\ \bibnamefont
			{White}},\ }\href {\doibase 10.1103/physrevlett.69.2863} {\bibfield
		{journal} {\bibinfo  {journal} {Physical Review Letters}\ }\textbf {\bibinfo
			{volume} {69}},\ \bibinfo {pages} {2863} (\bibinfo {year}
		{1992})}\BibitemShut {NoStop}%
	\bibitem [{\citenamefont {Schollwöck}(2011)}]{Schollwoeck2011}%
	\BibitemOpen
	\bibfield  {author} {\bibinfo {author} {\bibfnamefont {U.}~\bibnamefont
			{Schollwöck}},\ }\href {\doibase 10.1016/j.aop.2010.09.012} {\bibfield
		{journal} {\bibinfo  {journal} {Annals of Physics}\ }\textbf {\bibinfo
			{volume} {326}},\ \bibinfo {pages} {96} (\bibinfo {year} {2011})}\BibitemShut
	{NoStop}%
	\bibitem [{\citenamefont {Hubig}\ \emph {et~al.}(2015)\citenamefont {Hubig},
		\citenamefont {McCulloch}, \citenamefont {Schollwöck},\ and\ \citenamefont
		{Wolf}}]{Hubig2015}%
	\BibitemOpen
	\bibfield  {author} {\bibinfo {author} {\bibfnamefont {C.}~\bibnamefont
			{Hubig}}, \bibinfo {author} {\bibfnamefont {I.~P.}\ \bibnamefont
			{McCulloch}}, \bibinfo {author} {\bibfnamefont {U.}~\bibnamefont
			{Schollwöck}}, \ and\ \bibinfo {author} {\bibfnamefont {F.~A.}\ \bibnamefont
			{Wolf}},\ }\href {\doibase 10.1103/physrevb.91.155115} {\bibfield  {journal}
		{\bibinfo  {journal} {Physical Review B}\ }\textbf {\bibinfo {volume} {91}},\
		\bibinfo {pages} {155115} (\bibinfo {year} {2015})}\BibitemShut {NoStop}%
	\bibitem [{\citenamefont {Hubig}\ \emph {et~al.}()\citenamefont {Hubig},
		\citenamefont {Lachenmaier}, \citenamefont {Linden}, \citenamefont
		{Reinhard}, \citenamefont {Stenzel}, \citenamefont {Swoboda}, \citenamefont
		{Grundner},\ and\ \citenamefont {Mardazad}}]{HubigSyTen}%
	\BibitemOpen
	\bibfield  {author} {\bibinfo {author} {\bibfnamefont {C.}~\bibnamefont
			{Hubig}}, \bibinfo {author} {\bibfnamefont {F.}~\bibnamefont {Lachenmaier}},
		\bibinfo {author} {\bibfnamefont {N.-O.}\ \bibnamefont {Linden}}, \bibinfo
		{author} {\bibfnamefont {T.}~\bibnamefont {Reinhard}}, \bibinfo {author}
		{\bibfnamefont {L.}~\bibnamefont {Stenzel}}, \bibinfo {author} {\bibfnamefont
			{A.}~\bibnamefont {Swoboda}}, \bibinfo {author} {\bibfnamefont
			{M.}~\bibnamefont {Grundner}}, \ and\ \bibinfo {author} {\bibfnamefont
			{S.}~\bibnamefont {Mardazad}},\ }\href {https://syten.eu} {\enquote {\bibinfo
			{title} {The \textsc{SyTen} toolkit},}\ }\BibitemShut {NoStop}%
	\bibitem [{sup()}]{supp}%
	\BibitemOpen
	\bibinfo {note} {See Supplemental Material for additional data.}\BibitemShut
	{Stop}%
	\bibitem [{\citenamefont {Anderson}(1987)}]{Anderson1987}%
	\BibitemOpen
	\bibfield  {author} {\bibinfo {author} {\bibfnamefont {P.~W.}\ \bibnamefont
			{Anderson}},\ }\href {\doibase 10.1126/science.235.4793.1196} {\bibfield
		{journal} {\bibinfo  {journal} {Science}\ }\textbf {\bibinfo {volume}
			{235}},\ \bibinfo {pages} {1196} (\bibinfo {year} {1987})}\BibitemShut
	{NoStop}%
	\bibitem [{\citenamefont {Gutzwiller}(1963)}]{Gutzwiller1963}%
	\BibitemOpen
	\bibfield  {author} {\bibinfo {author} {\bibfnamefont {M.~C.}\ \bibnamefont
			{Gutzwiller}},\ }\href {\doibase 10.1103/PhysRevLett.10.159} {\bibfield
		{journal} {\bibinfo  {journal} {Physical Review Letters}\ }\textbf {\bibinfo
			{volume} {10}},\ \bibinfo {pages} {159} (\bibinfo {year} {1963})}\BibitemShut
	{NoStop}%
	\bibitem [{\citenamefont {Koepsell}\ \emph {et~al.}(2021)\citenamefont
		{Koepsell}, \citenamefont {Bourgund}, \citenamefont {Sompet}, \citenamefont
		{Hirthe}, \citenamefont {Bohrdt}, \citenamefont {Wang}, \citenamefont
		{Grusdt}, \citenamefont {Demler}, \citenamefont {Salomon}, \citenamefont
		{Gross},\ and\ \citenamefont {Bloch}}]{Koepsell2021}%
	\BibitemOpen
	\bibfield  {author} {\bibinfo {author} {\bibfnamefont {J.}~\bibnamefont
			{Koepsell}}, \bibinfo {author} {\bibfnamefont {D.}~\bibnamefont {Bourgund}},
		\bibinfo {author} {\bibfnamefont {P.}~\bibnamefont {Sompet}}, \bibinfo
		{author} {\bibfnamefont {S.}~\bibnamefont {Hirthe}}, \bibinfo {author}
		{\bibfnamefont {A.}~\bibnamefont {Bohrdt}}, \bibinfo {author} {\bibfnamefont
			{Y.}~\bibnamefont {Wang}}, \bibinfo {author} {\bibfnamefont {F.}~\bibnamefont
			{Grusdt}}, \bibinfo {author} {\bibfnamefont {E.}~\bibnamefont {Demler}},
		\bibinfo {author} {\bibfnamefont {G.}~\bibnamefont {Salomon}}, \bibinfo
		{author} {\bibfnamefont {C.}~\bibnamefont {Gross}}, \ and\ \bibinfo {author}
		{\bibfnamefont {I.}~\bibnamefont {Bloch}},\ }\href {\doibase
		10.1126/science.abe7165} {\bibfield  {journal} {\bibinfo  {journal}
			{Science}\ }\textbf {\bibinfo {volume} {374}},\ \bibinfo {pages} {82}
		(\bibinfo {year} {2021})}\BibitemShut {NoStop}%
	\bibitem [{\citenamefont {Tu}\ \emph {et~al.}(2018)\citenamefont {Tu},
		\citenamefont {Schindler}, \citenamefont {Neupert},\ and\ \citenamefont
		{Poilblanc}}]{Tu2018}%
	\BibitemOpen
	\bibfield  {author} {\bibinfo {author} {\bibfnamefont {W.-L.}\ \bibnamefont
			{Tu}}, \bibinfo {author} {\bibfnamefont {F.}~\bibnamefont {Schindler}},
		\bibinfo {author} {\bibfnamefont {T.}~\bibnamefont {Neupert}}, \ and\
		\bibinfo {author} {\bibfnamefont {D.}~\bibnamefont {Poilblanc}},\ }\href
	{\doibase 10.1103/physrevb.97.035154} {\bibfield  {journal} {\bibinfo
			{journal} {Physical Review B}\ }\textbf {\bibinfo {volume} {97}},\ \bibinfo
		{pages} {035154} (\bibinfo {year} {2018})}\BibitemShut {NoStop}%
	\bibitem [{\citenamefont {Szasz}\ \emph {et~al.}(2020)\citenamefont {Szasz},
		\citenamefont {Motruk}, \citenamefont {Zaletel},\ and\ \citenamefont
		{Moore}}]{Szasz2020}%
	\BibitemOpen
	\bibfield  {author} {\bibinfo {author} {\bibfnamefont {A.}~\bibnamefont
			{Szasz}}, \bibinfo {author} {\bibfnamefont {J.}~\bibnamefont {Motruk}},
		\bibinfo {author} {\bibfnamefont {M.~P.}\ \bibnamefont {Zaletel}}, \ and\
		\bibinfo {author} {\bibfnamefont {J.~E.}\ \bibnamefont {Moore}},\ }\href
	{\doibase 10.1103/physrevx.10.021042} {\bibfield  {journal} {\bibinfo
			{journal} {Physical Review X}\ }\textbf {\bibinfo {volume} {10}},\ \bibinfo
		{pages} {021042} (\bibinfo {year} {2020})}\BibitemShut {NoStop}%
	\bibitem [{\citenamefont {Kamilla}\ \emph {et~al.}(1996)\citenamefont
		{Kamilla}, \citenamefont {Wu},\ and\ \citenamefont {Jain}}]{Kamilla1996}%
	\BibitemOpen
	\bibfield  {author} {\bibinfo {author} {\bibfnamefont {R.~K.}\ \bibnamefont
			{Kamilla}}, \bibinfo {author} {\bibfnamefont {X.~G.}\ \bibnamefont {Wu}}, \
		and\ \bibinfo {author} {\bibfnamefont {J.~K.}\ \bibnamefont {Jain}},\ }\href
	{\doibase 10.1016/0038-1098(96)00126-3} {\bibfield  {journal} {\bibinfo
			{journal} {Solid State Communications}\ }\textbf {\bibinfo {volume} {99}},\
		\bibinfo {pages} {289} (\bibinfo {year} {1996})}\BibitemShut {NoStop}%
	\bibitem [{\citenamefont {W{\'{o}}js}\ and\ \citenamefont
		{Quinn}(2002)}]{Wojs2002}%
	\BibitemOpen
	\bibfield  {author} {\bibinfo {author} {\bibfnamefont {A.}~\bibnamefont
			{W{\'{o}}js}}\ and\ \bibinfo {author} {\bibfnamefont {J.~J.}\ \bibnamefont
			{Quinn}},\ }\href {\doibase 10.1103/physrevb.66.045323} {\bibfield  {journal}
		{\bibinfo  {journal} {Physical Review B}\ }\textbf {\bibinfo {volume} {66}},\
		\bibinfo {pages} {045323} (\bibinfo {year} {2002})}\BibitemShut {NoStop}%
	\bibitem [{\citenamefont {Doretto}\ \emph {et~al.}(2005)\citenamefont
		{Doretto}, \citenamefont {Goerbig}, \citenamefont {Lederer}, \citenamefont
		{Caldeira},\ and\ \citenamefont {Smith}}]{Doretto2005}%
	\BibitemOpen
	\bibfield  {author} {\bibinfo {author} {\bibfnamefont {R.~L.}\ \bibnamefont
			{Doretto}}, \bibinfo {author} {\bibfnamefont {M.~O.}\ \bibnamefont
			{Goerbig}}, \bibinfo {author} {\bibfnamefont {P.}~\bibnamefont {Lederer}},
		\bibinfo {author} {\bibfnamefont {A.~O.}\ \bibnamefont {Caldeira}}, \ and\
		\bibinfo {author} {\bibfnamefont {C.~M.}\ \bibnamefont {Smith}},\ }\href
	{\doibase 10.1103/physrevb.72.035341} {\bibfield  {journal} {\bibinfo
			{journal} {Physical Review B}\ }\textbf {\bibinfo {volume} {72}},\ \bibinfo
		{pages} {035341} (\bibinfo {year} {2005})}\BibitemShut {NoStop}%
	\bibitem [{\citenamefont {Balram}\ \emph {et~al.}(2015)\citenamefont {Balram},
		\citenamefont {Wurstbauer}, \citenamefont {W{\'{o}}js}, \citenamefont
		{Pinczuk},\ and\ \citenamefont {Jain}}]{Balram2015}%
	\BibitemOpen
	\bibfield  {author} {\bibinfo {author} {\bibfnamefont {A.~C.}\ \bibnamefont
			{Balram}}, \bibinfo {author} {\bibfnamefont {U.}~\bibnamefont {Wurstbauer}},
		\bibinfo {author} {\bibfnamefont {A.}~\bibnamefont {W{\'{o}}js}}, \bibinfo
		{author} {\bibfnamefont {A.}~\bibnamefont {Pinczuk}}, \ and\ \bibinfo
		{author} {\bibfnamefont {J.~K.}\ \bibnamefont {Jain}},\ }\href {\doibase
		10.1038/ncomms9981} {\bibfield  {journal} {\bibinfo  {journal} {Nature
				Communications}\ }\textbf {\bibinfo {volume} {6}},\ \bibinfo {pages} {8981}
		(\bibinfo {year} {2015})}\BibitemShut {NoStop}%
\end{thebibliography}
%

\newpage
\widetext
\newpage
\begin{center}
	\textbf{\Large{Supplemental Material}}
\end{center}
\setcounter{equation}{0}
\setcounter{figure}{0}
\setcounter{table}{0}
\setcounter{page}{1}
\makeatletter
\renewcommand{\theequation}{S\arabic{equation}}
\renewcommand{\thefigure}{S\arabic{figure}}
\renewcommand{\bibnumfmt}[1]{[S#1]}

\section{Results for $L_y=4$}
\begin{figure}[h!]
	\centering
	\includegraphics{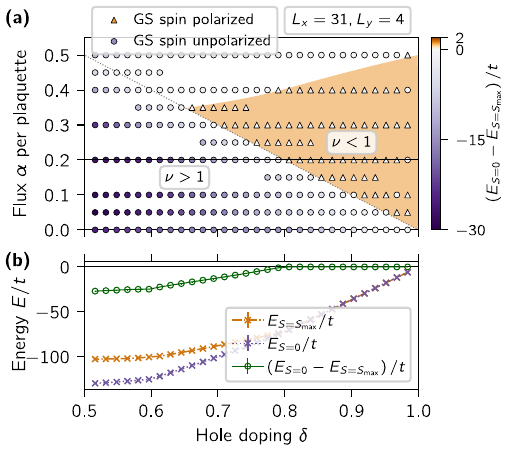}
	\caption{
		(a) Energy difference between the lowest energy states found in DMRG for the $S=S_{\rm max}$ and the $S=0$ sectors as function of magnetic flux per plaquette $\alpha$ and doping level $\delta$.
		The gray dotted line indicates $\nu=1$.
		For $\alpha \lesssim 0.35$ and $\nu\leq1$ (shaded region) the ground state is spin polarized with an almost degenerate spin-singlet excited state (see also (b)).
		In contrast, for $\nu > 1$ the spin-singlet is energetically favored significantly.
		At large flux, $\alpha \gtrsim \alpha_c \approx 0.4$, the QH ferromagnetism breaks down and we find the ground state to be unpolarized even for $\nu<1$.
		(b) Ground state energies in both sectors at $\alpha=0.2$ (solid line in (a)).
		Data is given for a system of size $L_x\times L_y=31\times 4$.
	}
	\label{fig:SuppMat:GSEnergies}
\end{figure}

\begin{figure}[h!]
	\centering
	\includegraphics{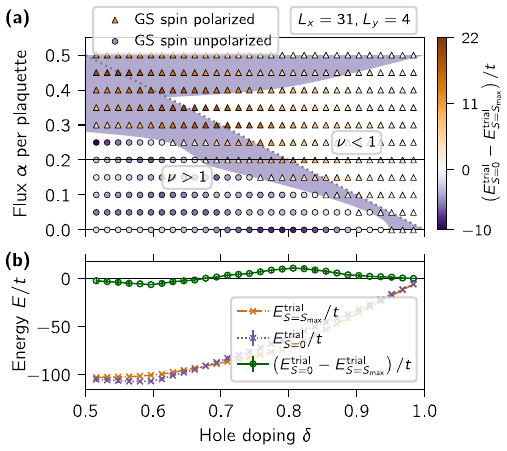}
	\caption{
		(a) Energy difference between the trial energies for the $S=S_{\rm max}$ and the $S=0$ sectors.
		The gray dotted line indicates $\nu = 1$ and the shaded area indicates the regime where the DMRG finds the ground state in the spin-singlet sector, while the trial states predict ferromagnetic order.
		(b) Trial state energies in both sectors at $\alpha=0.2$ (solid line in (a)).
		Data is given for a system of size $L_x\times L_y=33\times 5$.
	}
	\label{fig:SuppMat:GSTrialStates}
\end{figure}

To check that our results are not mere finite size effects, we also studied a system of size $L_x \times L_y = 31\times4$ in contrast to the larger system in the main text.
Again, we performed DMRG simulations (see Fig.~\ref{fig:SuppMat:GSEnergies}) as well as estimations of the ground state energy using the trial states discussed in the main text (see Fig.~\ref{fig:SuppMat:GSTrialStates}).
We find that the qualitative features discussed in the main text are visible here as well.
While we did not perform a systematic extrapolation to the thermodynamic limit, we believe the results to be valid there as well.

\section{Incompressibility of the QH Ferromagnet}
In the continuum, the QH~ferromagnet is known to be an incompressible state.
To verify that also the state found in our lattice studies shares this behavior, we calculate the energy cost $\mu_{N,+} = E_{N+1} - E_{N}$ to add a particle in the spin polarized state as function of the filling factor.
Ideally, for an incompressible state we expect this energy cost to have a discontinuity.
In the finite systems studied here, this discontinuity will be smeared out, but we still expect a sharp increase at $\nu=1$.

\begin{figure}[h!]
	\centering
	\includegraphics{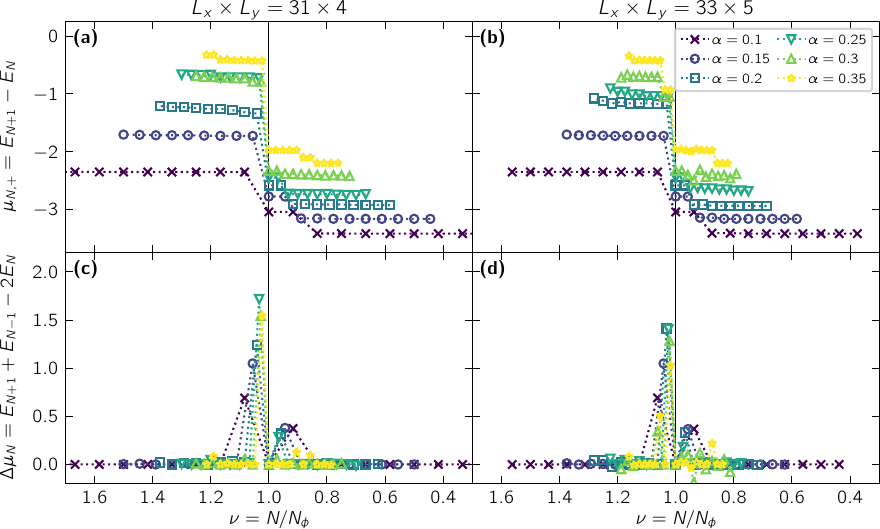}
	\caption{
		 (a,~b) Energy cost $\mu_{N,+}$ for adding a particle and (c,~d) charge gap $\Delta\mu_{N}$ in the spin polarized sector as function of the filling factor $\nu$ and varying flux $\alpha$.
		We find a clear jump of $\mu_{N,+}$ at $\nu=1$ indicating the incompressibility of the spin polarized state.
		Correspondingly, we observe a large charge gap $\Delta\mu_{\nu=1} \sim 0.5 t$ at this filling.
		This is in agreement with the spin polarized state  at $\nu=1$ being a lattice analog of the QH~ferromagnet.
		Data shown for systems of sizes $\L_x\times L_y = 31\times4$ (a,~c) and $33\times5$ (b,~d).
		\label{fig:SuppMat:Incompressibility}
	}
\end{figure}

Indeed, using our DMRG results for the energy in the spin polarized sector we find such a jump, see Fig.~\ref{fig:SuppMat:Incompressibility}(a,b).

To obtain the charge gap in this sector, we calculate $\Delta\mu_N = E_{N+1}+E_{N-1}-2E_N$.
We find a large charge gap $\Delta\mu_{\nu=1} \sim 0.5t$ at filling factor $\nu=1$, while the gap essentially closes away from this filling factor, see Fig.~\ref{fig:SuppMat:Incompressibility}(c,d).
This is another signature of the incompressible QH~ferromagnet at $\nu=1$.

\section{Spin Correlations around Quasi-Particles/Holes}
For the data points in the qh- and qp-skyrmion phases discussed in the main text, we analyzed the spin-spin correlations $C_{x_0}(x)$ relative to a reference site in the respective density structure as well as away from it.

\begin{figure}[h!]
	\centering
	\includegraphics{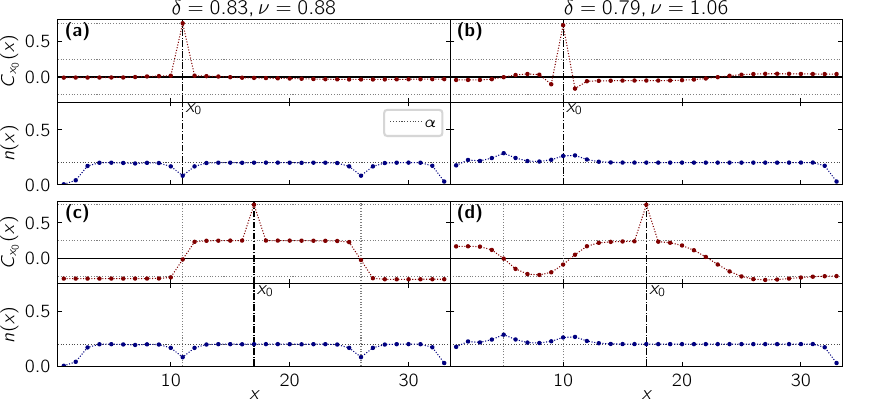}
	\caption{
		Spin-spin correlations $C_{x_0}(x)$ around (a) a quasi-hole and (b) a quasi-particle, with the reference site $x_0$ inside the density structure.
		(a) For the quasi-hole the spin texture is clearly in agreement with the expectation for a skyrmion.
		(b) Around the quasi-particle the situation is less clear, but some of the characteristic features remain.}
	\label{fig:SuppMat:qh/qp-skyrmion}
\end{figure}

We find that for the quasi-hole excitation the spin-spin correlations relative to the quasi-hole are clearly in agreement with the expectation for a skyrmion, see Fig.~\ref{fig:SuppMat:qh/qp-skyrmion}(a) and main text.
In particular, we find indication of local ferromagnetic correlations, while at larger distances from the quasi-hole the spins are anti-aligned.
We note, however, that the degree of spin (anti-)alignment is suppressed compared to the skyrmionic excitation at $\nu=1$.

Relative to a reference site in the center of the system, we find spin textures reminiscent of domain walls at the locations of the quasi-holes, see Fig.~\ref{fig:SuppMat:qh/qp-skyrmion}(c).

Around the quasi-particle the spins seem to be anti-aligned on a short length scale, see Fig.~\ref{fig:SuppMat:qh/qp-skyrmion}(b).
Nevertheless, at intermediate length scales, we find light signatures of spin alignment.
However, while the spin texture around the quasi-particle shares some features of the characteristic skyrmionic texture, the situation is less clear in this case.
We attribute this to the larger size of the quasi-particle and the Pauli correlation hole.

For a reference site in the center of the system, Fig.~\ref{fig:SuppMat:qh/qp-skyrmion}(d) and main text, we again find a behavior similar to that of domain walls related to the quasi-particles.
We note however, that compared to the quasi-hole case the domain walls are less sharp in this case and extend over a finite range.

\section{Spin Textures at Large Flux and Small Filling}

\begin{figure}
	\centering
	\includegraphics{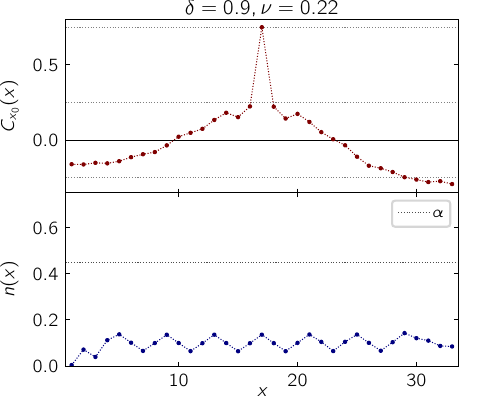}
	\caption{
		Spin-spin correlations $C_{x_0}(x)$ (upper panel) and local density $n(x)$ (lower panel) for large magnetic flux $\alpha=0.45 > \alpha_c$ and small filling factor, where the ground state is spin polarized.
		The overall behavior is reminiscent of the skyrmion textures discussed close to filling factor $\nu=1$ at smaller magnetic fields $\alpha \leq \alpha_c$.
		Data is given for a system of size $L_x\times L_y = 33 \times 5$ and $x_0 = 17$.
	}
	\label{fig:SuppMat:LargeAlpha}
\end{figure}

For large flux per plaquette, $\alpha > \alpha_c$, we found the ground state at $\nu=1$ to be a spin-singlet state.
However, upon reducing the particle number and hence the filling factor, we again find spin polarized ground states in the extremely dilute regime.
In particular, for some cases close to the transition between spin polarized and spin-singlet ground states we even observe spin correlations reminiscent of the skyrmion textures close to $\nu=1$ for $\alpha \lesssim \alpha_c$ discussed in the main text, see Fig.~\ref{fig:SuppMat:LargeAlpha}.

\end{document}